\documentclass[12pt]{iopart}
\usepackage{graphicx,subfig} 
\usepackage[format=plain,justification=RaggedRight,singlelinecheck=false,font=small,labelfont=bf,labelsep=space]{caption} 

\usepackage{color}

\makeatletter
\DeclareRobustCommand{\text}{%
  \ifmmode\expandafter\text@\else\expandafter\mbox\fi}
\let\nfss@text\text
\def\text@#1{{\mathchoice
  {\textdef@\displaystyle\f@size{#1}}%
  {\textdef@\textstyle\f@size{#1}}%
  {\textdef@\textstyle\sf@size{#1}}%
  {\textdef@\textstyle \ssf@size{#1}}%
  \check@mathfonts
  }%
}
\def\textdef@#1#2#3{\hbox{{%
                    \everymath{#1}%
                    \let\f@size#2\selectfont
                    #3}}}
\makeatother

\newcommand{\mum}{\ensuremath{{\mu}\text{m}}}

\begin{document}

\title[Deformation of flow-stabilized solids]{Nonaffine deformation under compression and decompression of a flow-stabilized solid}

\author{Carlos P. Ortiz, Robert Riehn and Karen E. Daniels}
\address{Department of Physics, North Carolina State University, Raleigh, NC, 27695, USA}

\vspace{10pt}
\begin{indented}
\item[] \today
\end{indented}

\begin{abstract}
Understanding the particle-scale transition from elastic deformation to plastic flow  is central to making predictions about the bulk material properties and response of disordered materials. 
To address this issue, we perform experiments on flow-stabilized solids composed of micron-scale spheres within a microfluidic channel, in a regime where particle inertia is negligible. 
Each solid heap exists within a stress gradient imposed by the flow, 
and we track the positions of particles in response to single impulses of fluid-driven compression or decompression.
We find that the resulting deformation field is well-decomposed into an affine field, with a constant strain profile throughout the solid, and a non-affine field. 
The magnitude of this non-affine response decays with the distance from the free surface in the long-time limit, suggesting that the distance from jamming plays a significant role in controlling the length scale of plastic flow. Finally, we observe that compressive pulses create more rearrangements than decompressive pulses, an effect that we quantify using  the  $D^2_\mathrm{min}$ statistic for non-affine motion. Unexpectedly, the time scale for the compression response is shorter than for decompression at the same strain (but unequal pressure), providing insight into the coupling between deformation and cage-breaking. 
\end{abstract}

\section{Introduction} 

Understanding how  structural rearrangements in disordered solids differ from crystalline solids is central \cite{Schall2007,Falk1998,Falk2011} to achieving control of material properties such as resistance to flow~\cite{Brady1993}, sound propagation ~\cite{Kriegs2004}, heat capacity~\cite{Lubchenko2007} , and dielectric constants~\cite{Bradshaw-Hajek2009}. For large deformations, the microscopic response differs non-perturbatively from the predictions of linear elasticity \cite{Bocquet2009}. Instead of linear deformations, phenomena such as shear banding \cite{Hays2000}, yielding and plastic rearrangements \cite{Hebraud1997}, and non-local effects~\cite{Lu2000} are present.  
Recent experiments have explored non-affine deformations in 
3D sheared colloidal glasses \cite{Chikkadi2012}, 3D emulsions~\cite{Knowlton2014}, and 2D foams~\cite{Twardos2005}. 
For sufficiently slow deformations, it is an open question whether  the flow behavior \cite{Lerner2012} is controlled by the jamming transition, where moduli vanish as the packing approaches a critical packing fraction \cite{Liu2010}.

\begin{figure}
\centering 
\includegraphics[width=0.8\textwidth]{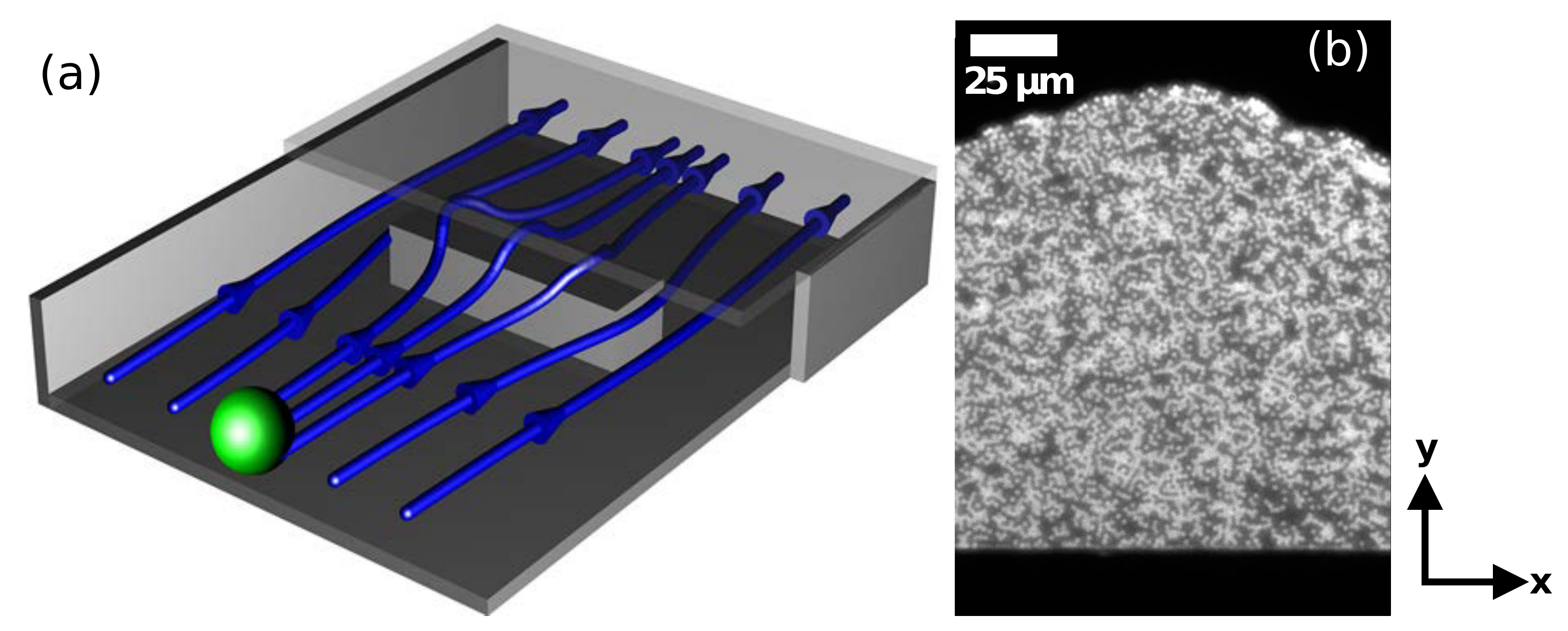}
\caption{(a) Schematic of channel geometry, not to scale, illustrating barrier with overflow. Streamlines are scaled versions of a calculation solving Stokes' equations on the true device dimensions.
(b) Image of central region of a heap of bidisperse particles. Dark regions are dimly fluorescent 600~nm particles and bright regions are brightly fluorescent 710~nm particles. White box indicates the location  of the region in which deformations are analyzed. \label{fig:setup}}
\end{figure}

In this paper, we present experiments quantifying the particle-scale deformation  of flow-stabilized solids: particle heaps formed under controlled hydrodynamic stress (see Fig.~\ref{fig:setup}). 
These quasi-2D heaps are assembled via the slow accumulation of micron-scale particles against a barrier within a microchannel, and are found to be stable above P\'eclet number~$\sim 1$ \cite{Ortiz2013}.  For lower P\'eclet numbers (slower flows) the particles reversibly evaporate away from the solid. We have previously observed that the elastic modulus of the solid is proportional to the confining stress provided by the fluid flow~\cite{Ortiz2014}. However,  the amount of  deformation of the solid in response to a flow perturbation is dependent on the sign of the perturbation: for piles prepared under identical conditions, compressions result in smaller strains than decompressions \cite{Ortiz2014}. At the bulk scale, this effect can be understood by considering an excluded volume equation of state, as in thermal systems, under the assumption of  locally affine deformations. 
In this paper, we investigate how the particle-scale dynamics lead to deviations from the excluded volume argument at high deformations.

Our experiments use sterically- and electrostatically-stabilized
Brownian microspheres with a short Debye length (3~nm), so that the net interparticle interaction is well-approximated by a hard-sphere potential except near contact. A  bidisperse mixture of particle sizes (5:4 diameter ratio) suppresses the nucleation of crystal domains.  Using fluorescence microscopy, we measure the particle-scale deformation fields and characterize the response of the heap under compressive and decompressive loads created by changing the hydrodynamic stress. We characterize the influence of cooperative motions by measuring the degree to which the  the deformation field locally deviates from global affine deformations. 

The affine (or homogeneous) component of the deformation field is the part that can be described by an affine transformation: rotation, shear, extension, or compression \cite{Wu2005}. After identifying the affine component of the deformation, the residual (or inhomogeneous) component is the non-affine deformation. For linearly-elastic materials, only affine deformations are present, but non-affine deformations can arise due to either thermally-driven cage-breaking events \cite{Weeks2002} or local rearrangements \cite{Falk1998}. Here, we quantify two effects: the total non-affine rearrangements, and the spatiotemporal dynamics of the response.  We observe, as expected \cite{Leonforte2006,Ellenbroek2009}, that non-affine deformation fields typically exhibit mesoscale correlations. Furthermore, the non-affine deformations are about twice as large for compressive deformations as compared with decompressive deformations of similar size, and happen over a shorter time scale. For both compressive and decompressive deformations, non-affine deformations continue after affine deformations have completed.

\section{Experimental Setup}

Our experiments begin by assembling a  microsphere heap by flowing a dilute suspension against a barrier (see Fig.~\ref{fig:setup}a). The microchannel is fabricated to have a height  $H = 897$~nm, higher than the height of a barrier ($h_b=694$~nm), so that the fluid overflow accumulates particles against the barrier of width $W=512~\mum$.  The heights are chosen to create a quasi-2D heap, shallow enough to suppress both stacked and non-stacked bilayer phases \cite{VanWinkle1986}. The suspension is pumped into the channel by compressing a reservoir at the inlet using a low pressure, piezoelectrically actuated, digital regulator (AirCom PRE1-UA1), at $P_0 = 10$~kPa above atmospheric pressure. After two hours of equilibration, the heap is $154~\mum$ deep ($30^\circ$ angle of repose) and contains approximately 40,000 particles, as shown in Fig.~\ref{fig:setup}b.  The coordinate system takes $\hat x$ parallel to the barrier and $\hat y$ perpendicular to the barrier, with the origin at center of the barrier; the fluid flow is in the $-\hat y$ direction.


The dilute, aqueous suspension is prepared at a concentration of  $\rho=180/(100~{\mu}\text{m})^2$ fluorescent  microspheres.  The particles are a bidisperse mixture of equal concentrations of  600~nm  and  710~nm  polystyrene microspheres ($\approx$6\% polydispersity, elastic modulus $4$~GPa from Bangs Laboratories). 
We use steric and electrostatic
stabilization (sulfate functionalized surface with $\zeta$-potential $=-60$~mV
and coated with Triton X-100) to provide reversible
inter-particle and channel-particle interactions.
The suspending fluid is a density-matched aqueous solvent at pH~5.4, buffered by citric acid    to prevent crystallization and with $17\%(w/v)$  sucrose to provide density-matching. The later suppresses segregation and sedimentation effects, important both at the barrier and at the inlet reservoir. Because the total particle brightness  scales approximately with the particle volume, and we are working near the diffraction limit, the  600~nm particles appear dimmer than the 710~nm particles. This effect aids in tracking the motion of the particles. 

We quantify the affine and non-affine deformation due to a pulse of either compression ($\Delta P > 0$) or decompression ($\Delta P < 0$). The decompression pulse results from a change from 10~kPa to 0.5~kPa over a time scale of 10~ms, and the compression pulse returns the pressure to 10~kPa. This corresponds to $\Delta P/P_0 = 19$ for the compression pulse and $\Delta P/P_0 = 0.95$ for the decompression pulse.
Imaging occurs in two phases. A full view of the initial heap is accessible with a $10\times$ objective, while experiments quantifying the particle motions require visualizing a zoomed-in region using a $60\times$ objective with a $4\times$ beam expander. The image was recorded by a CCD camera with $10 \times 10$~$\mu$m$^2$ pixels and an exposure time 36\,$\mu s$.
As shown in Fig.~\ref{fig:setup}b, the zoomed measurement region is of size $40\,d \times 30\,d$ and is located adjacent to the barrier. For each pulse, we  first acquire an image of a region of interest at the center of the heap prior to the pressure change, allowing us to extract the initial configuration of particle positions. Additional images, taken at 27~Hz, characterize the particle-scale response of the heap to the change in pressure. After a wait of 100 seconds, long enough for particles to settle onto a new, equilibrated configuration, we repeat this process for the compression pulse. 

\section{Image Processing \label{sec:imageproc}}

\begin{figure}
\centerline{\includegraphics[width=\textwidth]{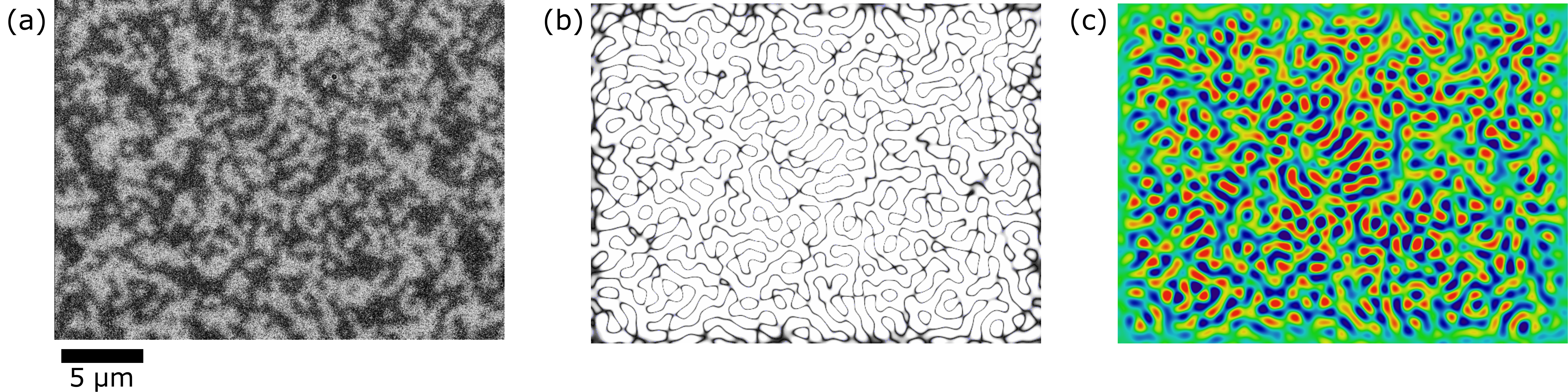}}
\caption{Illustration of image analysis process. (a) Sample micrograph, with grayscale indicating raw data. (b) Edge detection between bright and dark regions.
(c) Wiener Deconvolved Image. 
}
\label{fig:imanalysis}
\end{figure}

For either compressive or decompressive  pulses, we first compare the initial and final configurations (separated by 100 sec), and quantify both the  total deformation and the non-affine deformation. 
Second, using the series of frames immediately following the pulse, we track individual particles to identify non-affine effects on the local scale. Below, these are referred to as ``long-time'' and ``short-time'' dynamics, and require slightly different image-processing. For the long-time dynamics, the total distance traveled is on the order of a few particle diameters. Therefore, we first subtract the  total affine deformation before performing particle tracking using the Blair-Dufresne implementation \cite{MatlabTracking} of the Grier-Crocker particle tracking algorithm \cite{Crocker1996}. 

\paragraph{Particle identification:}
Fig.~\ref{fig:imanalysis} summarizes how we obtain particle positions beginning from a raw image. 
We identify the location of each particle by performing a Wiener deconvolution on the raw image, using a Gaussian approximation to the point-spread function with full-width at half maximum of 540~nm. 
This value is found to maximize the contrast in the output image,  as measured from the ratio of the standard deviation to the mean intensity, but is more effective at locating the large (bright) particles than the small (dim) particles.  
The resulting deconvolved image allows us to  detect the centroid of each particle using  Matlab's local extended maxima method.

\paragraph{Total deformation:}
We estimate the total affine deformation $\Delta y$ due to a single pulse by making a coarse-grained measurement of the particle displacements between an initial image and a final image. These two images are created by averaging 10 initial images $I_i(x,y)=\langle I_i(x,y,t)\rangle_t$ and 10 final images $I_f(x,y)=\langle I_f(x,y,t)\rangle_t$.
We divide $I_f(x,y)$ into horizontal strips of width $2d$ 
and compute a cross-correlation with $I_i(x,y)$ to determine its displacement. We find that the cross-correlation is sharply-peaked function for strips of at least this width.

Due to the large total strains, we perform particle pair matching between initial and final configurations based on particle positions from which the total affine deformation $\Delta y$ has already been subtracted. After this adjustment, pair identification proceeds as in the one-step particle tracking \cite{MatlabTracking}, with the size of the search region selected to correspond to the estimate of the maximum non-affine displacement amplitude, plus an estimate of the error in the affine strain. 

\paragraph{Short-time particle tracking:} In order to obtain particle trajectories during the full duration of the dynamics, we make several assumptions about the nature of valid trajectories. We limit the displacement per frame to $0.5$~\mum; this value is consistent with the total affine deformation rate determined above.
In addition, we consider a particle's identified size (brightness) in order to either split incorrect trajectories or or reconnect broken trajectories.

\section{Results}

In previous experiments \cite{Ortiz2014}, we observed that flow-stabilized solids exhibit a nonlinear stress-strain relationship in which the magnitude of the deformation of the {\it surface} of the flow-stabilized solid is well-described by 
\begin{equation}
\frac{\Delta y_\text{surface}}{y_\text{surface}} \propto \left(\frac{1}{1+\frac{\Delta P}{P_0}}-1\right).
 \label{eqn:gamma}
\end{equation}
The success of this description is somewhat surprising, as Eqn.~\ref{eqn:gamma} does not contain any information about the distribution of stresses or strains {\it throughout} the flow-stabilized solid.  The stress field within the solid  is anticipated to be similar to that in a sedimentation experiment where particles ``on top'' of the sedimented material apply some stress on lower layers (in the limit of shallow sediments without side walls). We believe that the success in describing our experiments is due to the universality of the van-der-Waals thermal argument.  However, that argument breaks down if non-affine motions occur, and we anticipate that the anticipated lower stress at the upstream (``top'') surface of the flow-stabilized solid is not fully characterized by the van-der-Waals argument.  In the following, we first identify the distributions of particle displacements in the asymptotic long-term limit, before following individual trajectories through compression and decompression. Our particular interest is in the associated particle-scale non-affine motions and their dependence on the sign of $\Delta P$.

\subsection{Total deformation \label{sec:longTdyn}}

\begin{figure}
\centerline{\includegraphics[width=5in]{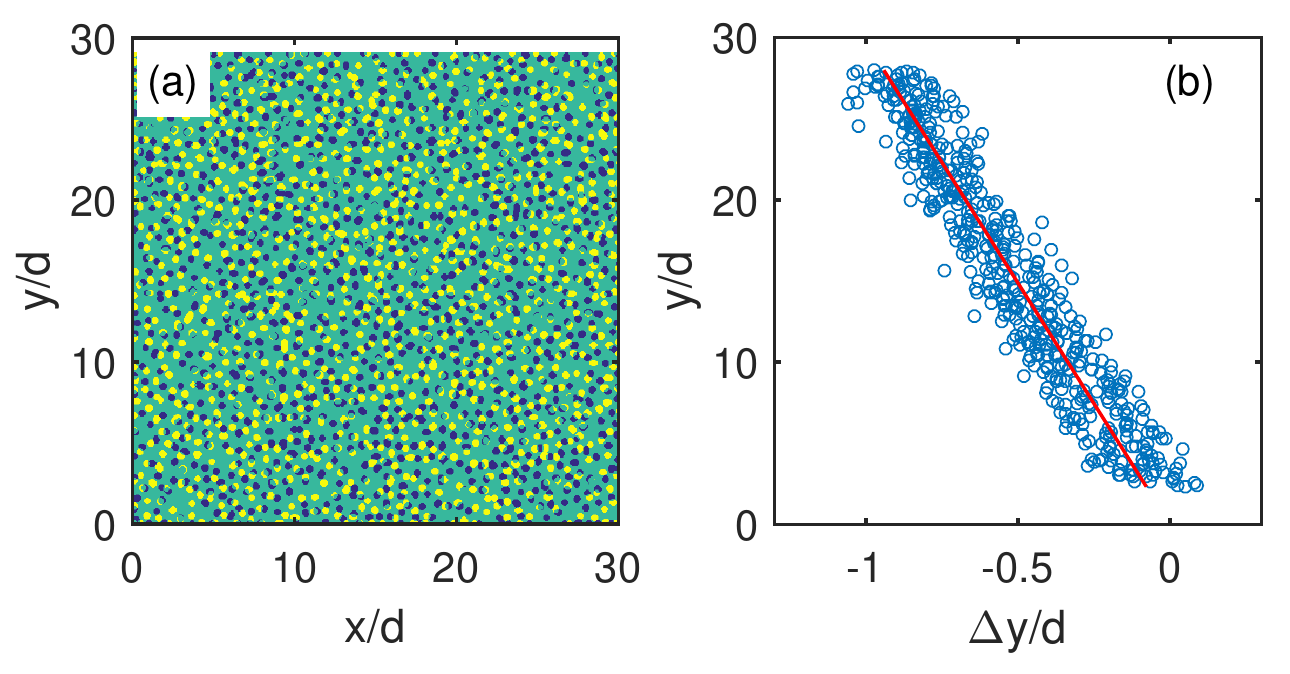}}
\caption{Determination of the total (long-time) affine field from particle tracks for $\Delta P=+9.5$~kPa.
(a) Image difference of blobs with size corresponding to $1\sigma$ of the position uncertainty in initial (blue) and final (yellow) configuration.  
(b) The associated affine deformation field.  The dashed line is a linear regression  with $\gamma = 3.4\pm 0.2$\%.
}
\label{fig:longTaffine}
\end{figure}

We find displacements of individual particles in the heap immediately before and 100~s after a compressive/decompressive pulse through a two-step analysis.   Following the homogeneous strain field assumption from our prior work \cite{Ortiz2014},  we first use the image cross-correlation analysis of images before and after deformation 
to obtain a global estimate of the affine strain field.
We then  use the affine transformation identified by the cross-correlation analysis as a scaffold for the matching of particles in the images before and after deformation.
In Fig.~\ref{fig:longTaffine}a, we show an example of the particle locations 
after Wiener deconvolution and prior to finding the centroids, for both $I_i$ (red, before compression) and $I_f$ (white, after compression).  By tracking each centroid, we can plot the local displacement $\Delta y$ as a function of  $y$-position  within the heap. 

As shown in In Fig.~\ref{fig:longTaffine}b, the mean behavior is linear,  confirming that the overall assumption of an affine deformation was sufficiently accurate. The best fit line to these points provides a measure of the strain:  $\Delta y = \gamma_\infty y$ with $\gamma_\infty=-3.8\%$. The precision of these measurements is insufficient to estimate the expected higher-order (quadratic) term, although we expect one to be present due to a depth-dependent stress field. The observed linear behavior, combined with Eqn.~\ref{eqn:gamma}, suggests that the packing fraction is close to invariant along the $y$-direction.


\begin{figure}
\centerline{\includegraphics[width=6in]{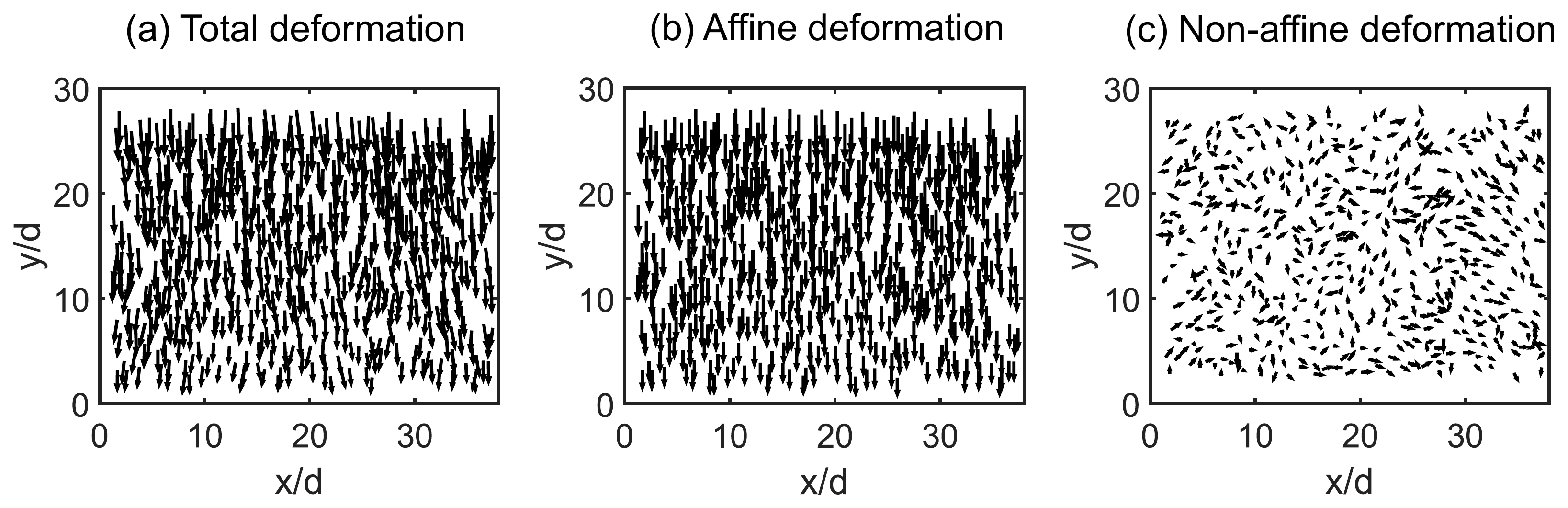}}
\caption{Deformation field for a compression given by a pressure change at the inlet of $\Delta P=+9.5$~kPa. The axes in both figures are the $x$-position and the $y$-position in units of particle diameters.  (a) Full deformation field. (b) Affine deformation field. (b) Non-affine deformation field, magnified by a factor of two.}
\label{fig:longTcompress}
\end{figure}

\begin{figure}
\centerline{\includegraphics[width=6in]{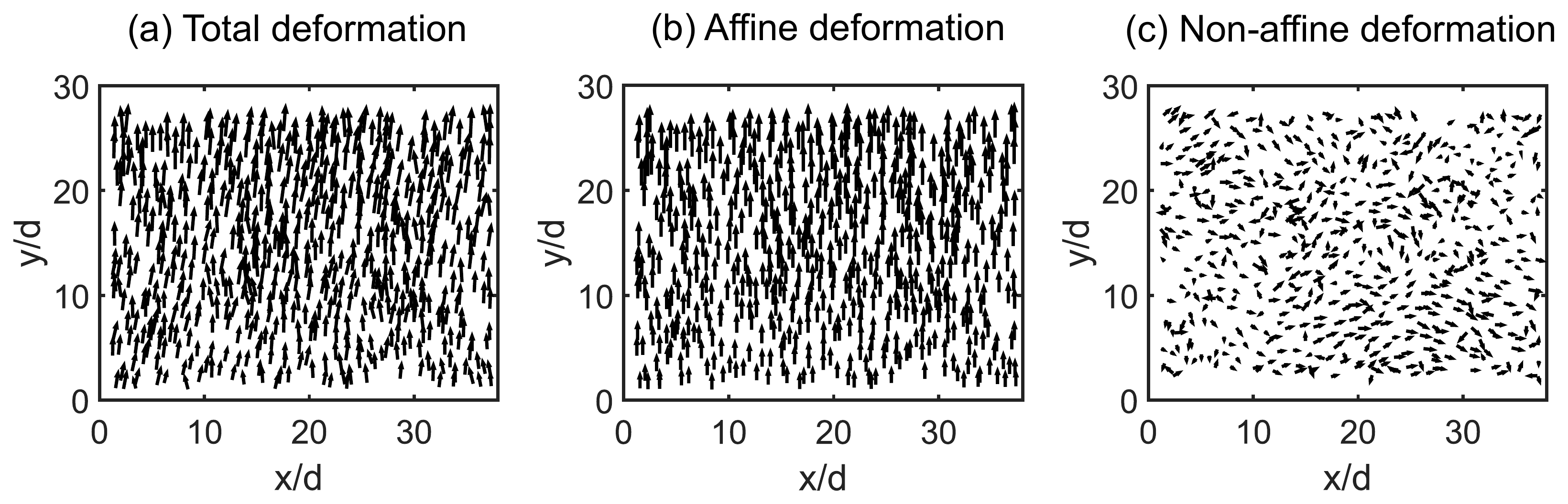}}
\caption{Deformation field for a decompression given by a pressure change at the inlet of $\Delta P=-9.5$~kPa. The axes in both figures are the $x$-position and the $y$-position in units of particle diameters.  (a) Full deformation field. (b) Affine deformation field. (b) Non-affine deformation field, magnified by a factor of two.}
\label{fig:longTdecompress}
\end{figure}

To obtain the total non-affine deformation field, we subtract the local affine motion from each displacement vector, as in \cite{Falk1998}. Figs.~\ref{fig:longTcompress} and \ref{fig:longTdecompress} (both compression and decompression) show the total, affine, and non-affine displacement fields, for comparison.  
Importantly, we observe bands of correlated motions, as expected from \cite{Leonforte2006,Ellenbroek2009}.
Because the total deformation field is not robust in tracking individual bead pairs over long times, we next examine the short-time dynamics.

\subsection{Dynamics of individual particle tracks \label{sec:shortTdyn} } 

\begin{figure}
\centerline{\includegraphics[width=5in]{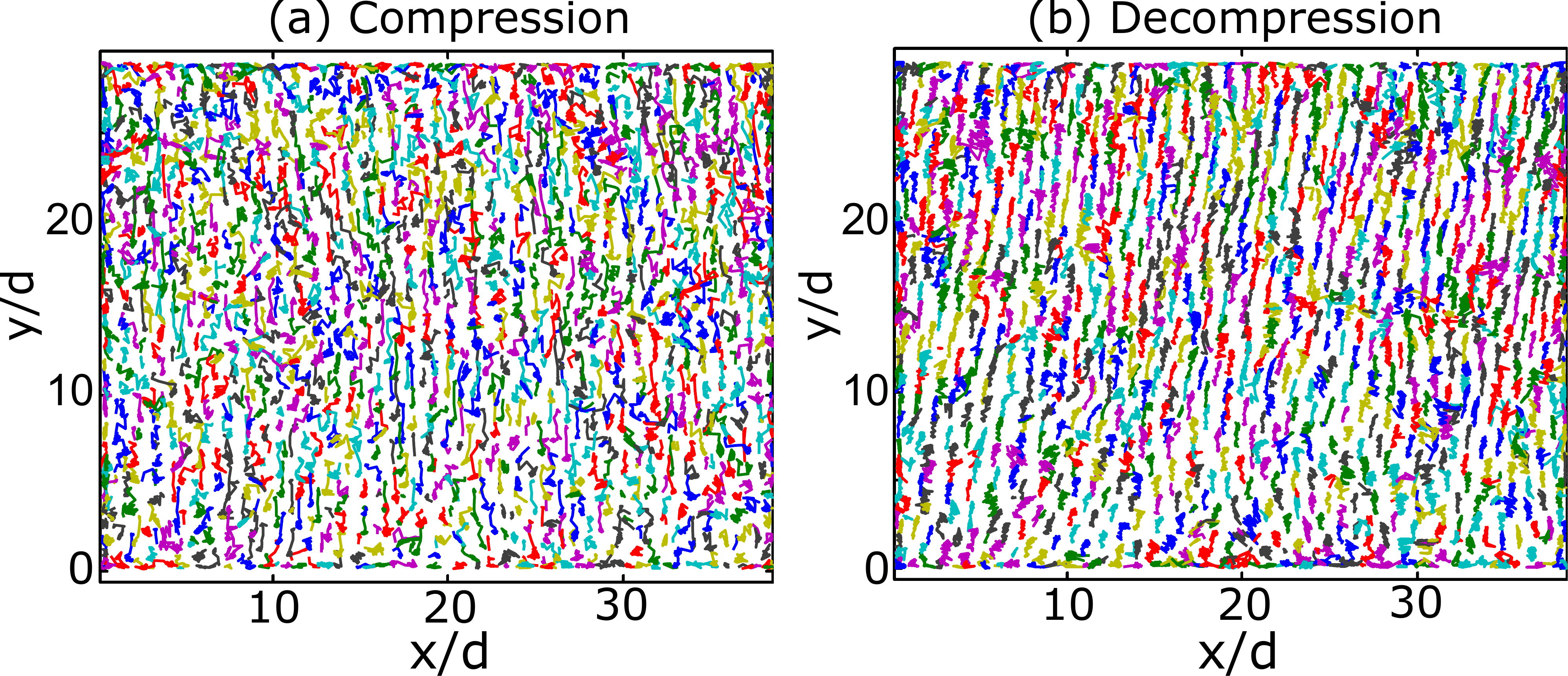}}
\caption{Particle Tracks and Packing Fraction. (a) Particle tracks during compression. Tracks are colored to introduce contrast. (b) Particle tracks during decompression. }
\label{fig:ptracks}
\end{figure}

Using the estimate of strain provided by Fig.~\ref{fig:longTaffine}b, we track the fast dynamics arising from a compressive/decompressive pulse.
For each frame, we first subtract the estimated affine deformation, based on the fraction of total strain which should have accumulated so far (see \S\ref{sec:imageproc}). This temporary adjustment allows for the correct association of particle centroids from frame to frame. Once the particle trajectories have been assembled based on these adjusted trajectories, we return to using the original positions detected for each particle centroid. The analyses that follow are based on the non-adjusted deformations that result from that tracking. 

Sample trajectories are shown in Fig.~\ref{fig:ptracks}. While the total deformation field is of the similar magnitude under compression and decompression, we find  
a more pronounced scrambling of the particle trajectories under compression, as compared to decompression. Below, we quantify both the affine and non-affine contributions to these trajectories.

\begin{figure}
\centerline{\includegraphics[width=4in]{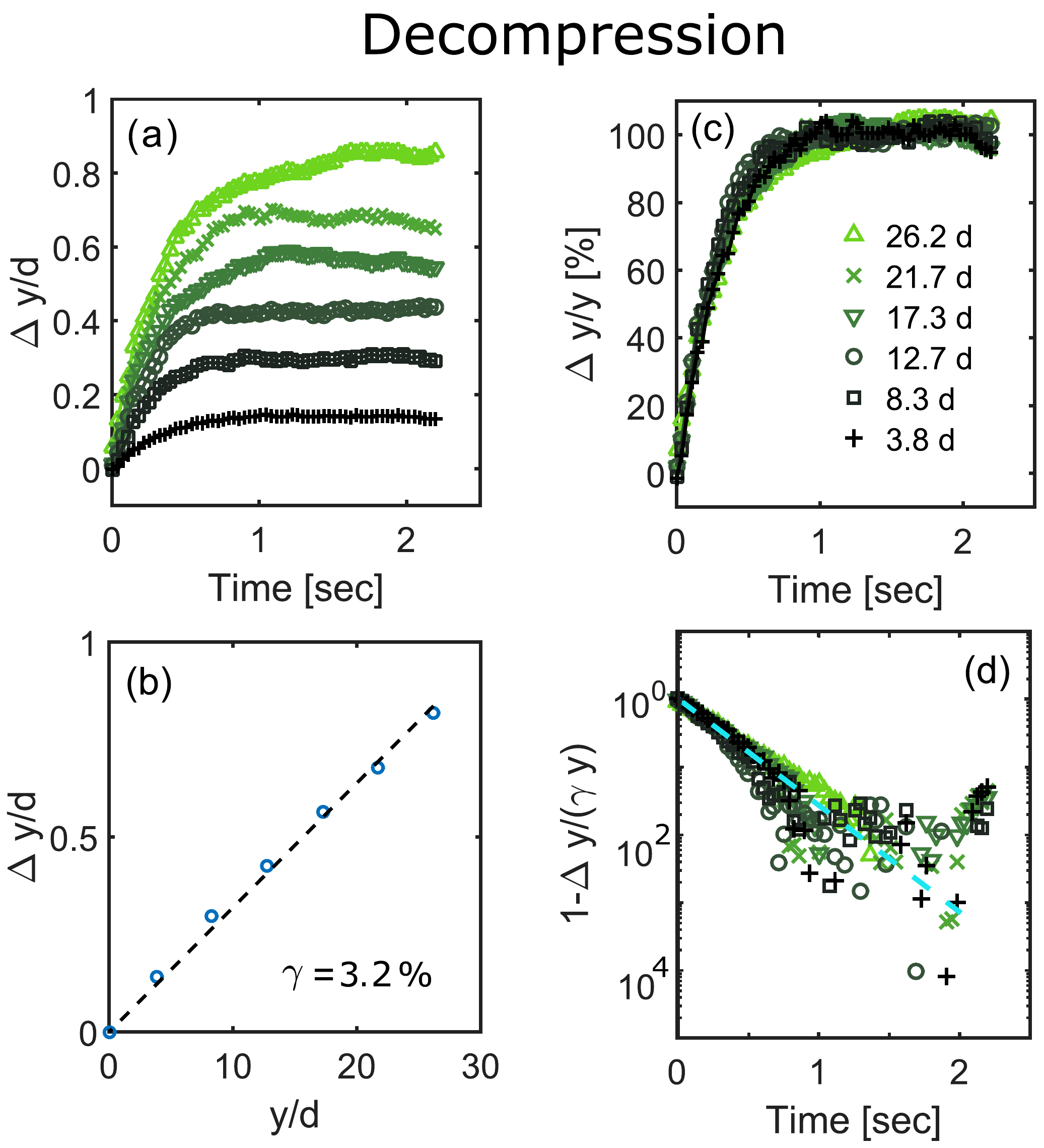}}
\caption{Affine Deformation Time-Series Analysis During Decompression. (a) Magnitude of correlated displacement field as a function of time at varying distances upstream of the barrier. (b) Long-time displacement amplitude $\Delta y$, as a function of distance upstream of the barrier, both in units of particle diameters. (c) Correlated displacement amplitude normalized by the long-time displacement amplitude. (d) Log-linear plot of growth curves in (c). }
\label{fig:shortT-decompress}
\end{figure}

\begin{figure}
\centerline{\includegraphics[width=4in]{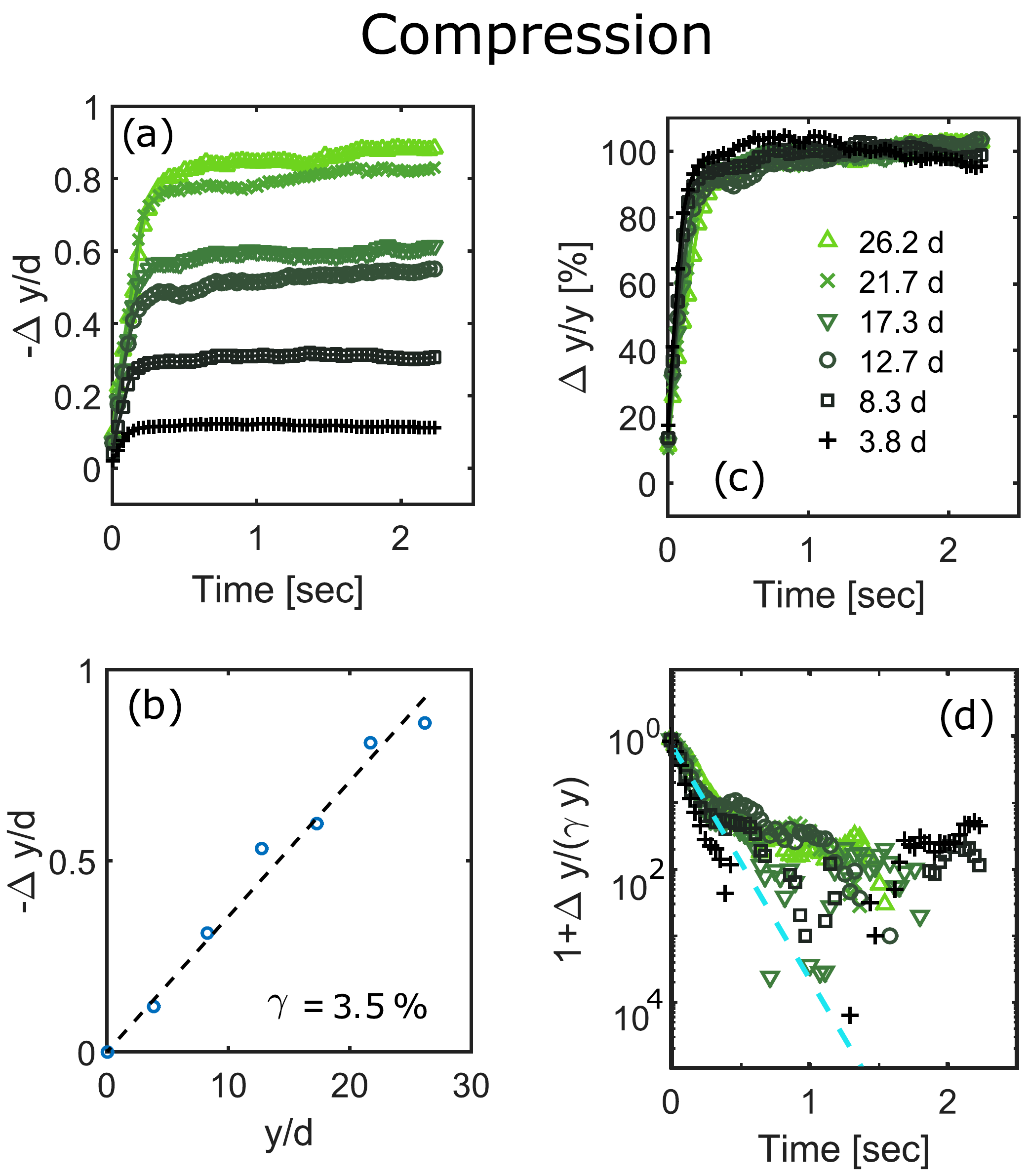}}
\caption{Affine Deformation Time-Series Analysis During Compression. (a) Magnitude of correlated displacement field as a function of time at varying $y$-positions. (b) Long-time displacement amplitude  $\Delta y$, as a function of distance upstream of the barrier, both in units of particle diameters. (c) Correlated displacement amplitude normalized by the long-time displacement amplitude. (d) Log-linear plot of growth curves in (c).  }
\label{fig:shortT-compress}
\end{figure}

\paragraph{Affine deformations: }
Data was binned within strips along the $x$-direction, providing ensembles of particles-dynamics sampled as a function of depth. Figs.~\ref{fig:shortT-decompress}a and \ref{fig:shortT-compress}a show the mean deformation field $\Delta y$ as a function of $y$-position for compression and decompression, respectively.  For both deformation directions, we find an exponential-like asymptotic approach to the final displacement magnitude. 
The depth-dependence of the asymptotic value of $\Delta y$ (Fig.~\ref{fig:shortT-decompress}b and \ref{fig:shortT-compress}b) demonstrates the same linear relationship originally shown in Fig.~\ref{fig:longTaffine}b.  The resulting slope ($\gamma_\infty = \Delta y/y$)  quantifies the dynamics of affine reorganization. We find a marked difference between decompression (Fig.~\ref{fig:shortT-decompress}c) and compression (Fig.~\ref{fig:shortT-compress}c) in that decompressions are far slower than compressions, and that the strain curves for decompression collapse better onto a single dynamic curve.

To quantify the difference, we make the Ansatz of a single-exponential approach to the asymptotic deformation
\begin{equation}\label{eqn:affinevt}
 \Delta y = \gamma_\infty \, y \left( 1-e ^{- \frac{\Delta t}{\tau}} \right)
\end{equation}
where $\Delta y$ is the  particle displacement after a time interval $\Delta t$,  
$\tau$ is a characteristic time scale of particle rearrangements, 
and $\gamma_\infty$ is the asymptotic strain. 
Note that the value $\gamma_\infty$ here is a fitting parameter; we find its value to be consistent with the estimate from the long-time dynamics.
As shown in both panels (d), this exponential form is a good fit for the decompression pulses with $\tau_\text{affine,decompression}=0.28\pm 0.05$~s.  
For compressive deformations, a single-exponential form is less consistent with the observed dynamics. Instead, there appears to be a two-step process of compression in which the viscous stress increase acts nearly instantaneously throughout the solid, while stresses due to particle-particle contacts propagate at a distinct speed of sound from the immobile barrier on which the solid is formed. 
Given the two-step nature of the process under compression, we establish an upper bound on the relaxation time scale of $\tau_\text{affine,compression}=0.11\pm 0.05$~s.

\begin{figure}
\centerline{\includegraphics[width=5in]{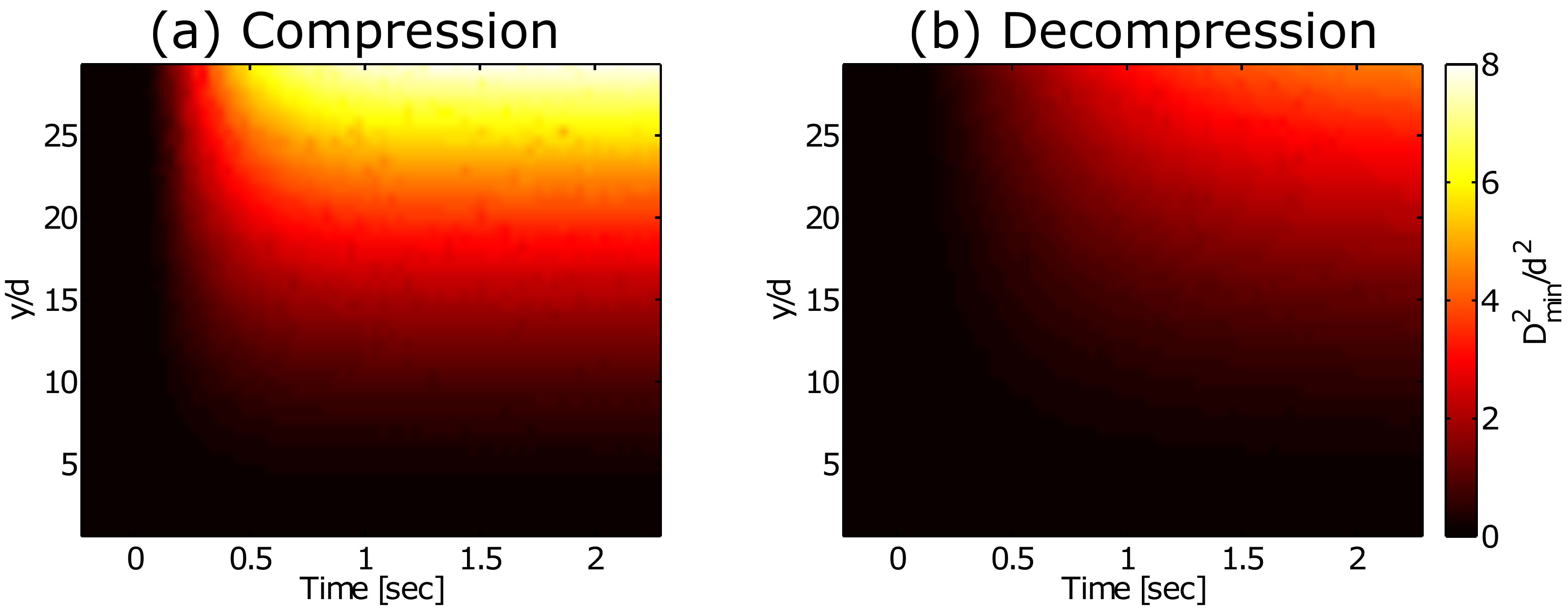}}
\caption{Magnitude of non-affine deformation field $D^2_\text{min}$ as a function of time and depth for (a) compression and (b) decompression.}
\label{fig:D2minim}
\end{figure}

\begin{figure} 
\centerline{\includegraphics[width=5in]{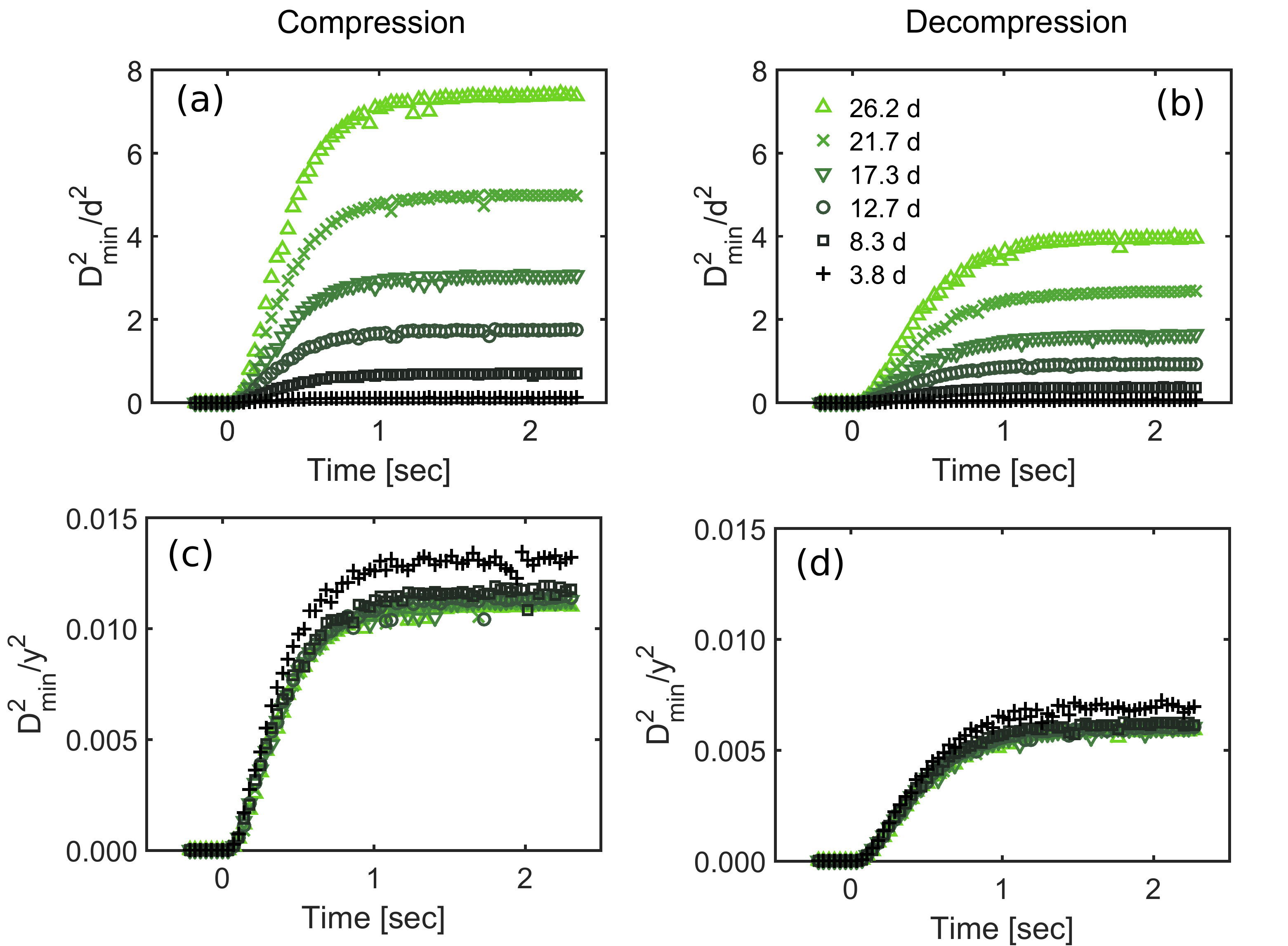}}
\caption{Plots of $D^2_\text{min}$ as a function of time, with the $y$-axis scaled either by the size of the large particles or the $y$-position at which the non-affine deformation is being probed. The legend in panel b applies to all panels. (a) Compression. $D^2_\text{min}$ scaled by $d^2$. (b) Compression. $D^2_\text{min}$ scaled by $y^2$. (c) Decompression. $D^2_\text{min}$ scaled by $d^2$. (d) Decompression. $D^2_\text{min}$ scaled by $y^2$.
\label{fig:D2minvt}}
\end{figure}

\paragraph{Non-affine deformations:} 
We identify the non-affine contribution to the deformation field by subtracting the affine portion associated with the best-fit instantaneous value of the strain, which we designate 
\begin{equation}
\gamma(t)=\gamma_\infty\left( 1-e ^{- \frac{\Delta t}{\tau}} \right).
\end{equation}
To quantify the resulting non-affinity, we use the $D_\text{min}^2$ measure \cite{Falk1998}  defined by 
\begin{equation}
D_\text{min}^2(t)\equiv\sum_\text{neighbors}\left(\Delta \vec r(t) - \gamma(t)\Delta \vec r(t_0)\right)^2 .
\end{equation} 
Here, $\vec r(t)$ is the set of local displacement vectors connection nearest neighbors, and $t_0$ is the time immediately before the pressure step was applied. Fig.~\ref{fig:D2minim} shows the time-evolution of the non-affine displacement as a function of $y$-position during compression and decompression, respectively. In both graphs, $D^2_\mathrm{min}$ grows and ultimately saturates. 
Interestingly, the magnitude of the non-affine field scales linearly with depth as demonstrated by the collapse of $D^2_\text{min}/y^2$ data series shown in Fig.~\ref{fig:D2minvt}.  This a surprising finding in light of the assumed constant strain throughout the flow-stabilized solid. Furthermore, the magnitude of non-affine deformations is approximately twice as large under compression than under decompression at near identical asymptotic strain $\gamma$.

We observe that the growth of $D^2_\text{min}$ with time is smooth. We are able to determine a characteristic time for the approach to the asymptotic value of $D^2_\text{min}$ by fitting a single exponential approach, as we did for the affine deformation field. In doing so, we neglect the low background value of $D^2_\text{min}$ in steady-state flow-stabilized solids arising from Brownian motion.  We find  $\tau_\text{non-affine,compression}=0.36\pm0.08$~s, and $\tau_\text{non-affine,decompression}=0.44\pm 0.1$~s.   Therefore, the non-affine field significantly lags the affine field for decompression ($\tau_\text{affine,decompression}=0.28\pm0.05$~s).  For compression, where a single time scale is less well defined, and an upper bound on the affine time scale is $\tau_\text{affine,compression}=0.11\pm 0.05$~s, the non-affine field also lags the affine deformation.  

\section{Discussion}

We have observed particle-scale non-affine motions within flow-stabilized solids, and examined how their spatiotemporal dynamics depend on whether the deformation is compressive or decompressive. We observed the typical swirling regions often associated with non-affine deformations, arising through  cooperatively rearranging regions. The magnitude of these effects is nearly twice as large under compression than under decompression, in spite of very similar total strains.

We observe that compressive pulses (large $\Delta P/P_0$) generate more non-affine deformation, which is able to dissipate the effect of the pulse more quickly. Because the non-affine fields for both compressive/decompressive deformations occur after similar delays with respect to the affine deformations, is suggests that they are triggered by the affine deformations. In the context of caging behavior, this suggests that the affine deformation distorts the cages provided by the neighboring particles and thereby makes Brownian cage-breaking (non-affine deformation) more likely.  Remarkably, this is the case even though the strain is approximately the same for decompression and compression.

In probing the spatial dependence to the magnitude of the non-affine deformations (Fig.~\ref{fig:D2minim}), we observe that the degree of non-affinity increases with distance from the barrier. This effect can be rescaled by the position to indicate a universal behavior. The form of this dependence suggests $D^2_\text{min}\propto \frac{1}{p^2} \propto \frac{1}{K}$, for pressure $p$ and modulus $K$ \cite{Ortiz2014}.  One interpretation is that the surface of the heap is less rigid (smaller $K$), and therefore more prone to undergoing non-affine deformations (higher $D^2_\text{min}$). 
Similar effects have been observed in numerical simulations \cite{Ellenbroek2009}, where increasingly non-affine displacements are present in proximity to unjamming.

The significance of the above conclusions to soft-matter particle assemblies is to reinforce the centrality of understanding non-affine rearrangements  to link bulk properties of the material, such as its modulus and global stability, to local properties about the typical particle geometry and rearrangement timescales.  Based on these results, this experimental setup opens the possibility to explore this connection, by studying multiple orders of magnitude of heap sizes, under dynamically tunable interaction potentials and heap geometry, maintaining the ability relate particle-scale rearrangement dynamics to bulk properties.   By doing so, it should be possible to 
determine length and time scales at which localized and collective rearrangements have the greatest impact on bulk properties, and shed light on the general mechanisms by which it is feasible to control the bulk properties of soft matter systems.

\section*{Acknowledgments} We are grateful for support from the National
Science Foundation through an NSF Graduate Fellowship, grants DMR-0644743, DMS-0968258, DMR-1121107, MRSEC/DMR-112090, and INSPIRE/EAR-1344280. Research was also supported by US Army Research Office--Division of Earth Materials and Processes grant 64455EV. This work was performed in part at the Cornell NanoScale Facility, a member of the National Nanotechnology Infrastructure Network, which is supported by the National Science Foundation (Grant ECCS-0335765).
This work was also performed in part at North Carolina State University facilities: Nanofabrication Facility, Advanced Instrumentation Facility, and Education and Research Laboratory.

\section*{References}

\bibliographystyle{iopart-num}

\providecommand{\newblock}{}

\end{document}